\documentclass[%
 aip,
cp,  
 amsmath,amssymb,
 reprint,%
]{revtex4-2}

\usepackage{graphicx}
\usepackage{dcolumn}
\usepackage{bm}

\usepackage[utf8]{inputenc}
\usepackage[T1]{fontenc}
\usepackage{mathptmx} 
\usepackage[version=4]{mhchem}
\usepackage{float}

\begin{document}

\title{Effect of Harvest on MTE Calculated by Single Step Process for Stochastic Population Model Under Allee Effect}

\author{Çağatay Eskin} 
 \email{cagatay.eskin@metu.edu.tr}
\affiliation{
  Middle East Technical University, Department of Physics, Ankara 06800, Turkey
}

\author{Özgür Gültekin}
 \email[Corresponding author: ]{gultekino@yahoo.com}

\affiliation{Mimar Sinan Fine Arts University, Deparment of Statistics, Istanbul 34427, Turkey}
\affiliation{Mimar Sinan Fine Arts University, Department of Mathematics, Istanbul 34427, Turkey}

\date{\today} 

\begin{abstract}
In this study, first we expand the cubic population model under Allee effect with a quadratic harvest function that represents harvest effect. Then, using four reaction equations representing the micro-interactions within the population under the influence of demographic noise and harvest, we obtain the mean field equations containing the effect of the harvest from the solution of the master equation. In this way, a relationship is established between the micro and macro parameters of the population. As a result, we calculate Mean Time to Extinction (MTE) by using WKB approximation for single step processes and observe the effect of harvesting.
\end{abstract}

\maketitle

\section{\label{sec:level1}Introduction}
In recent years, there has been much interest in investigating the different characteristics of the Allee effect in both ecology, evolutionary biology and genetics \cite{Levine2013,Qin2017,Sun2016, Kramer2010,Saakian2018,Schreiber2003, Cheptou,Davis2004}. Allee effect can be defined as positive correlation between population density and mean individual fitness \cite{Courchamp82}. If there is a critical value for the population size below which per capita growth rate of the population remains negative, this is strong Allee effect. Also, even though there is a positive correlation between population density and mean individual fitness, if such a critical population size value is not observed, this is weak Allee effect.\paragraph{}
There are studies that takes into account the effect of harvest on single species population models \cite{Sun2016,Hauser2006}. Effect of harvest on logistic model have been discussed in the book, \textit{Elements of Mathematical Ecology} \cite{kot2003}. However, effect of harvest has not been studied sufficiently in cubic models containing Allee effect. A simple population model containing the Allee effect is often represented by a cubic equation similar to the logistic equation. Populations in nature are under the influence of internal and external noise. Therefore, deterministic population models do not provide realistic outcomes. There are two general methods for a realistic description of the population, taking into account the effects of noise \cite{Yu2018,Assaf2017,Sancho1982,Smith2016,Nieddu2014}. One method is to investigate the evolution of the population under internal and external noise fluctuations with the Fokker-Planck equation. The other method is to obtain a mean field equation by solving the master equation with the transition rates representing the transition of the population from $n$ individuals to $n+r$ individuals and by describing the micro-interactions within the population with a representation similar to the chemical reaction representation.
These studies are generally taken forward to calculate MTE. Overview of calculation of MTE for single-step and multi-step processes using WKB approximation discussed \cite{Escudero2009,Assaf2010} . In this study, a population model under Allee effect, expanded by harvest, is examined with the solution of Master equation and effects of harvest on MTE were evaluated by calculating MTE based on one step approach by WKB method.

\section{\label{sec:level1}
Harvest Containing Cubic Model for a Population \\Under Allee Effect}
We start by writing the equation describing the cubic population model in the presence of Allee effect:

\begin{eqnarray}
\frac{dx}{dt}=-\frac{r}{K} x^3 + r \Big( 1+\frac{m}{K}  \Big) x^2 - rmx -(H_1x^2 + H_2x),
\end{eqnarray}
Here, $x$ is average population size at instance $t$, $r$ is intrinsic growth rate of the population, $K$ is carrying capacity and $m$ is Allee threshold for positive values. For the model to be biologically meaningful, parameters other than $m$ must be positive. To describe strong Allee effect, $m>0$ and to describe weak Allee effect $m<0$ must be taken. Also due to biological descriptions of parameters, $m<K$, and $H_1x^2 + H_2x$ describes harvest. When we take \\ $r^2(1 +m/K -H_1/r)^2>4r(rm+H_2)$ and $dx/dt =0$, $m^{'}$ and $K^{'}$ are the roots so that the equation is arranged as follows

\begin{eqnarray}
\frac{dx}{dt}= (rm+H_2)x \Big(1-\frac{x}{K^{'}} \Big) \Big(\frac{x}{m^{'}}-1 \Big)
\end{eqnarray}

\begin{eqnarray}
K^{'} = \frac{K}{2r} \bigg[r \Big(1+\frac{m}{K}-\frac{H_1}{r} \Big) + \sqrt{r^2 \Big(1 +\frac{m}{K}-\frac{H_1}{r} \Big)^2 -4\frac{r}{K}(rm+H_2)} \bigg] \\
m^{'} = \frac{K}{2r} \bigg[r \Big(1+\frac{m}{K}-\frac{H_1}{r} \Big) - \sqrt{r^2 \Big(1 +\frac{m}{K}-\frac{H_1}{r} \Big)^2 -4\frac{r}{K}(rm+H_2)} \bigg]
\end{eqnarray}
If $rm+H_2>0$, model represents strong Allee effect and if $rm+H_2<0$, shows weak Allee effect.

\section{\label{sec:level1}Stochastic Population Model}
We assume that the change in the average number of individuals over time is determined by Markovian birth and death processes. In order to write the mean field equations that will describe the time evolution of the average number of individuals in the population under the influence of demographic fluctuations, we write four reaction equations describing both the weak and strong Allee effect according to the value of the parameters, including harvest as follows:

  \begin{subequations}
    \begin{align}
      &X \ce{->[\mu]} (1+b-h_2)X\\
      &2X \ce{->[\lambda]} (2+a-h_1)X \\
      &X \ce{->[\gamma]} \varnothing \\
      &3X \ce{->[\beta]} (3-c)X
    \end{align}
  \end{subequations}
(5a) and (5c) represent linear birth and death processes expressed by $\mu$ and $\gamma$ constant multipliers, which do not include the effects of coexistence. The participation of additional mature offspring into the population with a $\lambda$ constant multiplier through binary cooperation is represented by (5b) and (5d) represents the death of individuals at a fixed $\beta$ rate as a result of the triple competition process for reasons such as overgrowth of the population in the presence of limited resources. Reaction (5d) is meaningful for $c=1,2,3$ values. Here $a$ and $b$ stand for the mature offspring and $h_1$ and $h_2$ stand for harvest. Including $h_2<b$ and $h_1<a$, we define $h_2$ as amount of harvest to reduce the number of newborns coming to reproductive age and $h_1$ as the amount of harvest to reduce additional offspring participation in reproductive age through binary cooperation. Time evolution of $P(n,t)$, probability for population to have $n$ individuals at instant $t$, is described by master equation which is linearized Chapman-Kolmogorov equation \cite{Gardiner2004Handbook}.

\begin{eqnarray}
\frac{dP(n,t)}{dt} = \sum_{r} \Big[W(n-r,r),P(n-r,t)-W(n,r)P(n,t) \Big]
\end{eqnarray}
Here, $W(n,r)$ defined as the rate of transition from the state where the population has $n$ number of individuals to the state where the number of individuals are $n+r$ .Transition rates written from (5) as follows \cite{Gardiner2004Handbook,Mendez2019a}:

\begin{subequations}
    \begin{align}
    &W( {n,a - {h_1}}) = \frac{\lambda }{2}n\left( {n - 1} \right)\\
    &W( {n,b - {h_2}}) = \mu n\\
    &W( {n, - 1} ) = \gamma n\\
    &W( {n, - c}) = \frac{\beta }{6}n\left( {n - 1} \right)\left( {n - 2} \right)
    \end{align}
\end{subequations}
When transition rates are put into master equation and each side multiplied by $n^k$ then summed up for all values of $n$ 

\begin{eqnarray}
\frac{\partial }{{\partial t}}\sum\limits_{n = 0}^\infty  {{n^k}} P\left( {n,t} \right) =&& \frac{\lambda }{2}\sum\limits_{n = 0}^\infty  {\left[ {{{\left( {n + a + {h_1}} \right)}^k} - {n^k}} \right]} n\left( {n - 1} \right)P\left( {n,t} \right)\nonumber\\
&&+ \mu \sum\limits_{n = 0}^\infty  {\left[ {{{\left( {n + b + {h_2}} \right)}^k} - {n^k}} \right]} nP\left( {n,t} \right)\nonumber\\
&&+ \gamma \sum\limits_{n = 0}^\infty  {\left[ {{{\left( {n - 1} \right)}^k} - {n^k}} \right]} nP\left( {n,t} \right)\nonumber\\
&&+ \frac{\beta }{6}\sum\limits_{n = 0}^\infty  {\left[ {{{\left( {n - c} \right)}^k} - {n^k}} \right]} n\left( {n - 1} \right)\left( {n - 2} \right),
\end{eqnarray}
is obtained. Thus, by using the definiton $\left\langle {{n^k}} \right\rangle  = \sum\limits_{n = 0}^\infty  {{n^k}P\left( {n,t} \right)} $ and taking $\left\langle {{n^k}} \right\rangle  = {\left\langle n \right\rangle ^k}$, $k=1$. Also, using mean field approximation and including $\left\langle n \right\rangle  = x$, mean field equation is found as follows:

\begin{eqnarray}
\frac{{dx}}{{dt}} = \left[ {\mu \left( {b - {h_2}} \right) - \gamma } \right]x + \frac{\lambda }{2}\left({a - {h_1}} \right){x^2} - \frac{\beta }{6}c{x^3}
\end{eqnarray}
Thus, when $dx/dt = 0$ is taken, roots of the equation (9) written as:

\begin{eqnarray}
m' = \frac{{3\lambda }}{{2c\beta }}\left[ {\left( {a - {h_1}} \right) - \sqrt {{{\left( {a - {h_1}} \right)}^2} + 8c\beta \frac{{\left[ {\mu \left( {b - {h_2}} \right) - \gamma } \right]}}{{3{\lambda ^2}}}} } \right]\\
K' = \frac{{3\lambda }}{{2c\beta }}\left[ {\left( {a - {h_1}} \right) + \sqrt {{{\left( {a - {h_1}} \right)}^2} + 8c\beta \frac{{\left[ {\mu \left( {b - {h_2}} \right) - \gamma } \right]}}{{3{\lambda ^2}}}} } \right]
\end{eqnarray}
For roots to be $N(1-\delta)$ and $N(1+\delta)$,

\begin{align}
N = \frac{{3\lambda \left( {a - {h_1}} \right)}}{{2\beta c}}&&
{\delta ^2} = 1 + \frac{{8\beta c\left[ {\mu \left( {b - {h_2}} \right) - \gamma } \right]}}{{3{\lambda ^2}{{\left( {a - {h_1}} \right)}^2}}}&&
{R_0} = \frac{{\mu \left( {b - {h_2}} \right)}}{\gamma }
\end{align}

we make this simplification. Thus, because of the condition on equation (2), for $\gamma-\mu(b-h_2)>0$ strong Allee effect, and for $\gamma-\mu(b-h_2)<0$ weak Allee effect is observed. We take time as $t \to \gamma t$ and population number density as $q = n/N$. For $q=O(1)$, $w_{r_i}(q)$ and $u_{r_i}(q)$ are $O(1)$. By using $W(Nq,r_i)=Nw_{r_1}(q)+u_{r_1}(q)+O(N^{-1})$ and $N \gg 1$, we rescale transition rates by applying general procedure \cite{Assaf2017}.

\begin{align}
{w_{{r_1}}} &= \frac{{2\left( {{R_0} - 1} \right)}}{{\left( {{\delta ^2} - 1} \right)\left( {a - {h_1}} \right)}}{q^2}           & {u_{{r_1}}} &=  - \frac{{2\left( {{R_0} - 1} \right)}}{{\left( {{\delta ^2} - 1} \right)\left( {a - {h_1}} \right)}}q \nonumber\\
{w_{{r_2}}} &= \frac{{{R_0}}}{{\left( {b - {h_2}} \right)}}q & {u_{{r_2}}} &= 0 \\
{w_{{r_3}}} &= q   &  {u_{{r_3}}}& = 0 \nonumber\\
{w_{{r_4}}} &= \frac{{{R_0} - 1}}{{\left( {{\delta ^2} - 1} \right)c}}{q^3} & {u_{{r_4}}} &=  - 3\frac{{{R_0} - 1}}{{\left( {{\delta ^2} - 1} \right)c}}{q^2} \nonumber
\end{align}

For single step case, meaning $a=b=c=1$, transition rates are calculated by ${w_ + }(q) = {w_{{r_1}}} + {w_{{r_2}}}$, ${w_ - }(q) = {w_{{r_3}}} + {w_{{r_4}}}$ and correction terms are calculated by ${u_ + }(q) = {u_{{r_1}}} + {u_{{r_2}}}$ , ${u_ - }(q) = {u_{{r_3}}} + {u_{{r_4}}}$

\section{\label{sec:level1}Effect of Harvest on MTE for Single Step Case Under Weak and Strong Allee Effect}
\begin{figure}[h]
\includegraphics[scale=0.6]{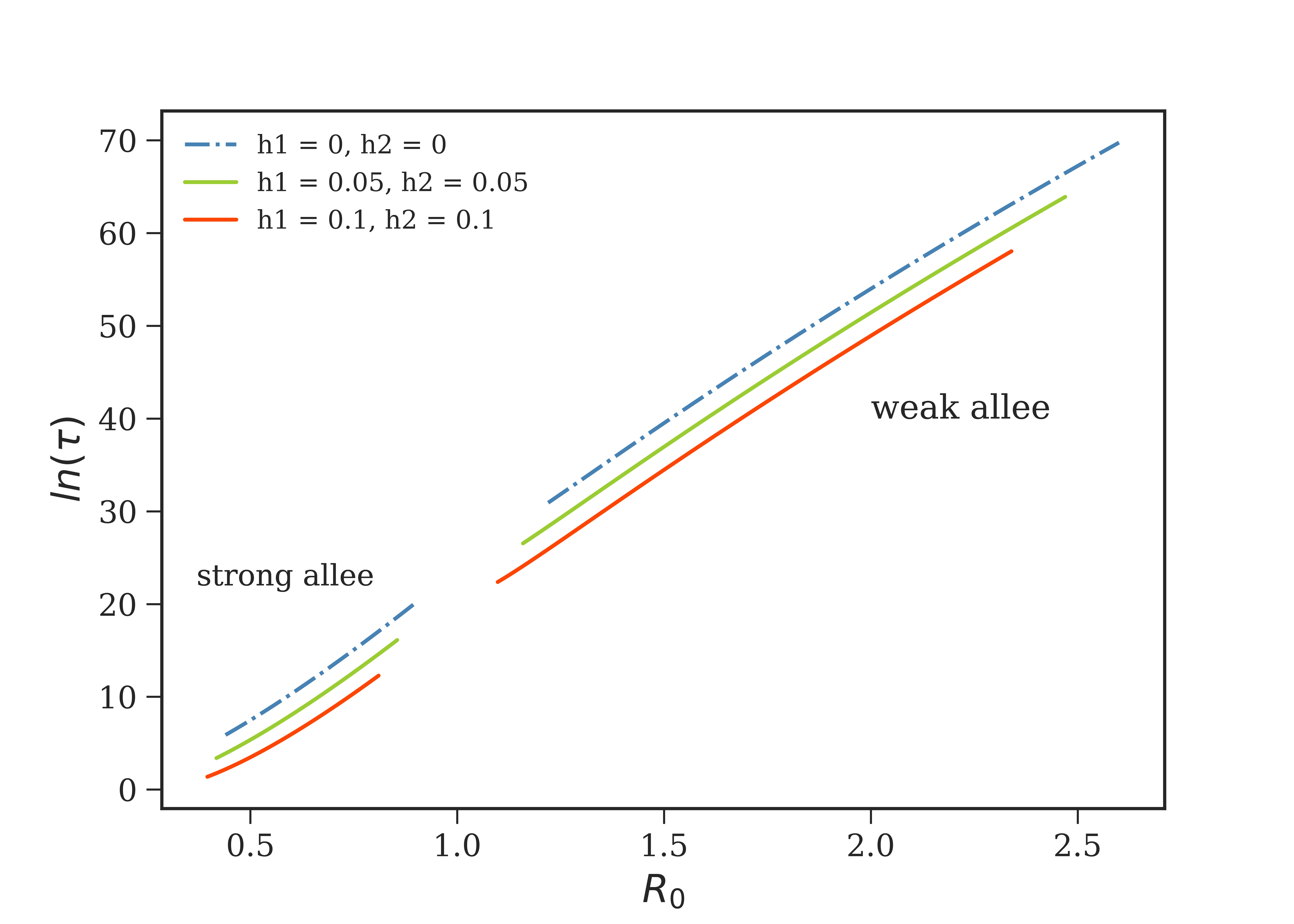}
\caption{\label{fig:plot} Change of $ln(\tau)$ with respect to $R_0$ for different values of $h_1$ and $h_2$.
Results are given for $\gamma=17.5, \lambda=1.5, \beta=0.05$.}
\end{figure}

The effect of harvest on MTE will be examined using single step process for strong and weak Allee effect cases. For $\gamma  - \mu (b - {h_2}) > 0$, meaning ${R_0} < 1$, strong Allee effect and for $\gamma  - \mu (b - {h_2}) < 0$, meaning ${R_0} > 1$, weak Allee effect, is observed. When $R_0 = 1$ WKB approximation does not work. For strong and weak Allee effect cases, MTE is calculated by equations (15) and (16), respectively \cite{Mendez2019a}. Comprehensive researches are available for these equations, where the MTE is calculated in a single step case with the WKB approach \cite{Assaf2017,Assaf2010}.
\begin{eqnarray}
\tau  =&& \frac{1}{{\gamma \left( {{R_0} - 1} \right)}}{\left[ {\frac{{2\pi {R_0}{{w'_ -} }\left( {q_e^s} \right)}}{{N{{w'_ +} }\left( {q_e^s} \right)\left[ {{{w'_ - }}\left( {q_e^s} \right){w_ + }\left( {q_e^s} \right) - {{w'_ +} }\left( {q_e^s} \right){w_ - }\left( {q_e^s} \right)} \right]}}} \right]^{0.5}}\\
&&\times{exp \Bigg[{N\int\limits_0^{q_e^s} {\ln \left[ {\frac{{{w_ + }\left( q \right)}}{{{w_ - }\left( q \right)}}} \right]} dq + \int\limits_0^{q_e^s} {\left[ {\frac{{{u_ + }\left( q \right)}}{{{w_ + }\left( q \right)}} - \frac{{{u_ - }\left( q \right)}}{{{w_ - }\left( q \right)}}} \right]} dq}}\Bigg]\nonumber
\end{eqnarray}
\begin{eqnarray}
\tau  =&& \frac{{2\pi }}{\gamma }{\left[ {\frac{{{w_ - }\left( {q_e^u} \right){w_ + }\left( {q_e^u} \right)}}{{\left[ {{{w'_ -} }\left( {q_e^s} \right){w_ + }\left( {q_e^s} \right) - {{w'}_ + }\left( {q_e^s} \right){w_ - }\left( {q_e^s} \right)} \right]\left| {{{w'_ -} }\left( {q_e^u} \right){w_ + }\left( {q_e^u} \right) - {{w'_ +} }\left( {q_e^u} \right){w_ - }\left( {q_e^u} \right)} \right|}}} \right]^{0.5}} \\
&&\times{exp \Bigg[{N\int\limits_{q_e^u}^{q_e^s} {\ln \left[ {\frac{{{w_ + }\left( q \right)}}{{{w_ - }\left( q \right)}}} \right]} dq + \int\limits_{q_e^u}^{q_e^s} {\left[ {\frac{{{u_ + }\left( q \right)}}{{{w_ + }\left( q \right)}} - \frac{{{u_ - }\left( q \right)}}{{{w_ - }\left( q \right)}}} \right]} dq}} \Bigg]\nonumber
\end{eqnarray}
The parameters used to calculate MTE in the previous section include the harvest effect. Single step situation to stochastically examine the harvesting effect gives only a limited idea. As can be seen from Fig.~\ref{fig:plot}, in the presence of harvest, MTE reduces. Meaning, population extinct in less time compared to the case where there is no harvest. As expected, if harvest gets bigger, MTE gets smaller. For a more realistic and comprehensive review of the harvesting effect on MTE, we recommend extending the research to a multi step situation.

\section{Conclusion}

The theory of first passage time or mean time to extinction has a wide range of applications, particularly in physics, chemistry and biology. We have proposed a stochastic model to examine the effect of harvest on time that takes for average size of population to reach an equilibrium point of extinction from another equilibrium point, under Allee effect and demographic fluctuations. We can say that the WKB approach we carried out with the single step process only has a limited meaning on average for the small values of $h_1=h_2$. Our analysis with the deterministic model predicts a limit in which weak Allee effect will turn into strong Allee effect in the presence of harvest. We think that utilizing the WKB approach with multi step processes to investigate how this transition boundary is affected in the presence of harvest will provide more valid and broader results.

\begin{acknowledgments}
We gratefully acknowledge Esra Yazıcıoğlu for her contributions.
\end{acknowledgments}

\nocite{*}
\bibliography{articles_used.bib}

\end{document}